\documentclass[fleqn,12pt,twoside]{article}
\usepackage{espcrc1}

\usepackage{graphicx}
\usepackage{graphics}
\usepackage[figuresright]{rotating}

\newcommand{\AmS}{{\protect\the\textfont2
  A\kern-.1667em\lower.5ex\hbox{M}\kern-.125emS}}
\title{Nucleons and nuclei in the context of low-energy QCD}

\author{Wolfram Weise\address[MCSD]{Physik-Department, Technische Universit\"at M\"unchen\\ 
D-85747 Garching, Germany}
\thanks{Work supported in part by BMBF and GSI}}
\begin{document}
\maketitle

\begin{abstract}
This presentation reports on recent developments concerning basic aspects of low-energy QCD as they relate to the understanding of the nucleon mass and the nuclear many-body problem. 
\end{abstract}

\section{INTRODUCTION: SOME BASICS OF LOW-ENERGY QCD}

Nuclear physics has a well developed phenomenology in terms of nucleons interacting
through potentials constrained by nucleon-nucleon scattering data. At the same time,  
nucleons and nuclei are aggregates of quarks and gluons residing in the hadronic,
low-temperature phase of QCD. In recent years, the understanding of the nucleon itself, in particular of
its low-energy, long-wavelength structure, has been progressing rapidly. Likewise, steps have been
taken to explore the possible connection of the nuclear many-body problem with low-energy QCD,
its symmetries and symmetry breaking patterns. It is therefore timely and appropriate to discuss these two related themes from a unifying perspective.

\subsection{Spontaneous chiral symmetry breaking and the hadronic mass gap}

At low energies confinement implies that the QCD eigenstates are not elementary quarks and gluons but hadrons, colour singlet composites of quarks and gluons. 
Confinement also implies that the chiral $SU(2)_L\times SU(2)_R$ symmetry of two-flavour QCD with (almost) massless 
$u$- and $d$-quarks is spontaneously broken. As a consequence, the QCD vacuum is non-trivial: it hosts a strong condensate of (scalar) quark-antiquark pairs, the chiral condensate $\langle \bar{q}q\rangle \simeq - 1.5$\,fm$^{-3}$. Goldstone's theorem states that there exist low frequency collective excitations of the condensed QCD ground state: pseudoscalar Goldstone bosons, identified with the pions.  Their mass vanishes in the exact chiral limit, i.e. for massless $u$- and $d$-quarks. The finite but small physical pion mass, $m_\pi \simeq 0.14\,$GeV, reflects the non-zero quark masses $m_{u,d} \sim 5\,$MeV. Goldstone's theorem also states that these pions interact weakly at low energy and momentum. In fact, their interaction strictly vanishes in the exact chiral limit and at zero energy and momentum.

A charged pion decays through weak interaction into a muon and an antineutrino. The hadronic matrix element of the axial vector current connecting the pion with the QCD vacuum determines the pion decay constant, $f_\pi$, with the empirical value $f_\pi = 92.4\pm 0.3\,$MeV.  This constant reflects, as an order parameter, the spontaneously broken chiral symmetry of the QCD ground state just like the chiral (quark) condensate itself. The symmetry breaking pattern of low-energy QCD is manifest in a characteristic mass gap that governs the spectrum of the lightest hadrons. The scale associated with this mass gap is $\Delta \sim 4\pi f_\pi \sim 1\,$GeV.

Low-energy QCD is the physics of strong interactions at energies and momenta small compared to the 1 GeV scale of spontaneous chiral symmetry breaking. In this low-frequency, long-wavelength limit QCD is realized in the form of an effective field theory of weakly interacting Goldstone bosons coupled to heavier hadrons which act as almost static sources. Given this scenario, we address the following questions:
\begin{itemize}
\item{How does the nucleon develop its mass, starting from almost massless quarks, and what role does the pion cloud of the nucleon play in this context?}
\item{What is the impact of spontaneously broken chiral symmetry in QCD on the nuclear many-body problem?}
\end{itemize}

\subsection{Chiral effective field theory}

The mass scale set by the gap $\Delta \sim 4 \pi f_{\pi}$ offers a
natural separation between "light" and "heavy" (or, correspondingly, "fast" and
"slow") degrees of freedom. The basic idea of an effective field theory is to
introduce the active light particles as collective degrees of freedom,  while the
heavy particles are treated as almost static. In low-energy QCD, the "light" particles are the pions whereas the nucleons figure as "heavy". The low-energy strong interaction dynamics is described by an effective Lagrangian which incorporates all relevant
symmetries of QCD. The elementary quarks and gluons of the original Lagrangian are now replaced by the Goldstone bosons of spontaneously broken chiral symmetry. 
Heavy particles, such as nucleons, act as sources for the Goldstone bosons. Their leading-order couplings are determined by the Noether currents of the underlying symmetries. 

In the sector of QCD with the two lightest quark flavours, relevant to most of nuclear physics, the Goldstone bosons are represented by a $2 \times 2$ matrix field $U (x) \in
SU(2)$ which collects the three isospin components $\pi_a (x)$ of the pion. A
convenient representation is $U (x) = \exp[i \tau_a \pi_a(x)/f]$ where the pion decay constant $f$ in the chiral limit provides a suitable normalisation. 
In essence, the QCD Lagrangian is replaced by an effective Lagrangian expressed in terms of $U(x)$ and its derivatives:
\begin{equation}
{\cal L}_{QCD} \rightarrow {\cal L}_{eff}(U,\,\partial U,\,\partial^2 U,\, ...)\,.
\end{equation}
Goldstone bosons do not interact unless they have non-zero four-momentum, so the low-energy expansion of ${\cal L}_{eff}$ is an ordering in powers of $\partial_\mu U$.

The effective Lagrangian in the pure meson sector (with zero baryon number) has the following
leading terms \cite{GL}:
\begin{equation}
{\cal L}_{eff}  = {\cal L}_{\pi} = {f^2\over 4}Tr[\partial_\mu U^\dagger\partial^\mu U]+ {f^2\over 2}B\, Tr[m(U^\dagger + U)] + {\cal O}[(\partial U)^4]\,.
\end{equation}
The second term on the r.h.s. represents the (perturbative) explicit chiral symmetry breaking by the quark mass matrix, $m = diag(m_u, m_d)$. The constant $B$ is related to the chiral condensate as $Bf^2 = -\langle\bar{q}q\rangle = m_\pi^2 f^2/(m_u + m_d)$. The last equality follows from identifying the pion mass term in the Lagrangian. With inclusion of higher order corrections the systematic expansion of the S matrix based on  ${\cal L}_{\pi}$, Chiral Perturbation Theory (ChPT), has become a powerful tool to deal with low-energy observables. One of its prime successes is the quantitative description of pion-pion scattering close to threshold \cite{CGL}.

Consider next the sector with one unit of baryon number which includes
the physics of the interacting pion-nucleon system. We restrict ourselves to the case of $N_f = 2$ flavours. 
The previous pure meson Lagrangian
is now replaced by ${\cal L}_{eff} = {\cal L}_{\pi} + {\cal L}_{\pi N}$. The additional term involving the 
nucleon is expanded again in powers of derivatives (external pion momenta) and quark masses.
The $\pi N$ effective Lagrangian to second order in the pion field has the form
\begin{eqnarray}
{\cal L}_{\pi N} & = & \bar{\Psi}_N(i\gamma_{\mu}\partial^{\mu} 
- M_N)\Psi_N - \frac{g_A}{2f_{\pi}} \bar{\Psi}_N\gamma_{\mu}\gamma_5 
\mbox{\boldmath $\tau$}\Psi_N\cdot\partial^{\mu}\mbox{\boldmath $\pi$}  \\
                   &   & -\frac{1}{4f_{\pi}^2}
 \bar{\Psi}_N\gamma_{\mu} 
\mbox{\boldmath $\tau$}\Psi_N\cdot\mbox{\boldmath $\pi$}\times
\partial^{\mu}\mbox{\boldmath $\pi$}
+ \frac{\sigma_N}{f_{\pi}^2} \bar{\Psi}_N\Psi_N\mbox{\boldmath $\pi$}^2 
+ ...~~.\nonumber
\end{eqnarray}
The Dirac spinor field $\Psi_N = (p,n)^T$ of the nucleon is represented
as an isospin-$1/2$ doublet of proton and neutron. The nucleon mass 
\begin{eqnarray}
M_N = M_0 + \sigma_N~~, 
\end{eqnarray}
has a large part $M_0$ which exists already in the chiral limit of vanishing quark masses. The sigma term
\begin{eqnarray}
\sigma_N = \langle N| m_q(\bar{u}u + \bar{d}d)|N\rangle 
\end{eqnarray}
is the correction to $M_0$ from the small quark mass term of the QCD Lagrangian, with $m_q = (m_u + m_d)/2$. Its empirical value is $\sigma_N \simeq 45 - 55$ MeV \cite{GLS}. The vector and axial vector couplings of the nucleon to the Goldstone boson fields are dictated by chiral symmetry. They involve known structure constants: the pion decay constant $f_\pi$ and the nucleon axial vector coupling constant, $g_A = 1.270 \pm 0.003$, determined from neutron beta decay.  
Not shown in Eq.(3) is a series of additional terms of order 
$(\partial^{\mu} \pi)^2$. These terms come with further constants that need to be fitted to experimental
data and reflect, for instance, the very important role played by the $\Delta(1232)$ resonance in pion-nucleon scattering. 

Baryon ChPT, the calculational framework based on ${\cal L}_{eff} = {\cal L}_{\pi} + {\cal L}_{\pi N}$, 
has been applied quite successfully to a variety of low-energy processes
(such as pion-nucleon scattering, threshold pion photo- and electroproduction and Compton scattering on the nucleon) for which increasingly accurate experimental data have become
available in the last decade. Reviews can be found in \cite{BDW}. In recent years the nucleon-nucleon interaction at long and intermediate distance scales has also become an active area of ChPT \cite{KBW,M}. 

\section{THE NUCLEON: ITS MASS AND SCALAR FIELD}

\subsection{Nucleon mass and pion cloud}

Understanding the nucleon mass is clearly one of the most fundamental issues in nuclear and particle physics \cite{TW}.  Progress is now being made towards a synthesis of lattice QCD and chiral effective field theory, such that extrapolations of lattice results to actual observables are beginning to be feasible \cite{LTY,PHW}. Accurate computations of the nucleon mass on the lattice have become available \cite{CPP,JLQ,QSF}, but so far with $u$ and $d$ quark masses exceeding their commonly accepted small values by typically an order of magnitude. Methods based on chiral effective theory can then be used, within limits, to interpolate between lattice results and physical observables.

The nucleon mass is determined by the expectation value $\langle N| \Theta_\mu^\mu |N\rangle$ of the trace of the QCD energy-momentum tensor, $\Theta_\mu^\mu = (\beta(g)/2g)G_{\mu\nu}G^{\mu\nu} + m_u\bar{u}u + m_d\,\bar{d}d + ...\,\,$, where $G^{\mu\nu}$ is the gluonic field tensor, $\beta(g)$ is the beta function of QCD, and $m_q\,\bar{q}q$ with $q = u,d,\,$ ... are the quark mass terms (omitting here the anomalous dimension of the mass operator for brevity). Neglecting small contributions from heavy quarks, the nucleon mass (4) taken in the $SU(2)_f$ chiral limit, $m_{u,d}\rightarrow 0$, is
\begin{equation}
M_0 = \langle N|{\beta\over 2g}G_{\mu\nu}G^{\mu\nu}|N\rangle ~.
\end{equation}
This relation emphasises the gluonic origin of the bulk part of $M_N$, the part for which lattice QCD provides an approriate tool to generate the relevant gluon field configurations. At the same time, QCD sum rules connect $M_0$ to the chiral condensate $\langle \bar{q}q\rangle$. The leading-order result (Ioffe's formula \cite{Io}) is:
\begin{equation}
M_0 = -{8\pi^2\over \Lambda_B^2}\langle \bar{q}q \rangle  + ... ~,
\end {equation}
where $\Lambda_B \sim 1$ GeV is an auxiliary scale (the Borel mass) which separates "short" and "long" distance physics in the QCD sum rule analysis. While Ioffe's formula needs to be improved by including condensates of higher dimension, it nevertheless demonstrates the close connection between dynamical mass generation and spontaneous chiral symmetry breaking in QCD. For nuclear physics,
Eq.(7) gives important hints \cite{WW}: the change of the condensate $\langle \bar{q}q \rangle$ with increasing baryon density implies a significant reduction of the nucleon mass in the nuclear medium. This observation finds its correspondence in the strong scalar field as it appears in phenomenological relativistic mean field models \cite{SW,RF}. 

In chiral effective field theory, the quark mass dependence of $M_N$ translates into a dependence on the pion mass, $m_\pi^2 \sim m_q$, at leading order. The dressing of the nucleon with its pion cloud, at one-loop order, is illustrated in Fig.1. The systematic chiral expansion of the nucleon mass gives an expression of the form \cite{PHW}:
\begin{equation}
M_N = M_0 +  c m_\pi^2 + d m_\pi^4 - {3\pi \over 2} g_A^2 m_\pi\left({m_\pi \over 4\pi f_\pi}\right)^2 \left(1-{m_\pi^2 \over 8M_0^2}\right) + {\cal O}(m_\pi^6)~,
\end{equation}
where the coefficients $c$ and $d$ multiplying even powers of the pion mass include low-energy constants constrained by pion-nucleon scattering. The coefficient $d$ also involves a $log\, m_\pi$ term.
Note that the piece of order $m_\pi^3$ (non-analytic in the quark mass) is given model-independently in terms of the known weak decay constants $g_A$ and $f_\pi$ (strictly speaking: by their values in the chiral limit). 

The interpolation shown in Fig.2 determines the nucleon mass in the chiral limit, $M_0$, and the sigma term $\sigma_N = \sum_{q=u,d} m_q(dM_N/dm_q)$. One finds $M_0 \simeq 0.89$ GeV and $\sigma_N = (47 \pm 3)$ MeV \cite{PHW} in this approach. (Treating the $\Delta$ isobar as an explicit degree of freedom rather than absorbing it in low-energy constants, there is a tendency for slightly larger values of $\sigma_N$).
 
\begin{figure}[htb]
\begin{minipage}[t]{70mm}
\includegraphics[scale=0.25,angle=0]{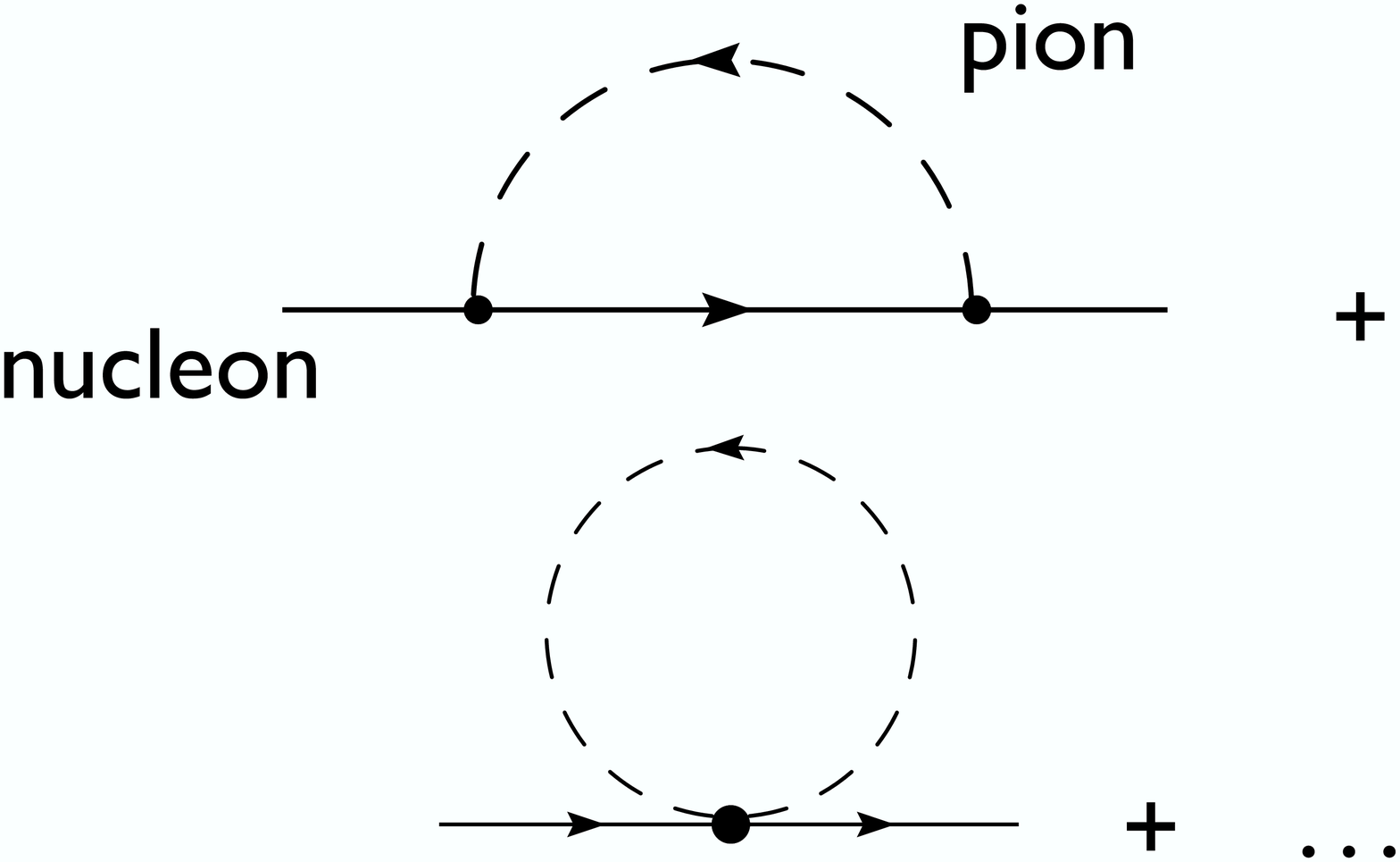}
\caption{Pion cloud contributions to the nucleon mass generated by the chiral effective Lagrangian (3).}
\label{fig.1}
\end{minipage}
\hspace{\fill}
\begin{minipage}[t]{85mm}
\includegraphics[scale=0.550,angle=0]{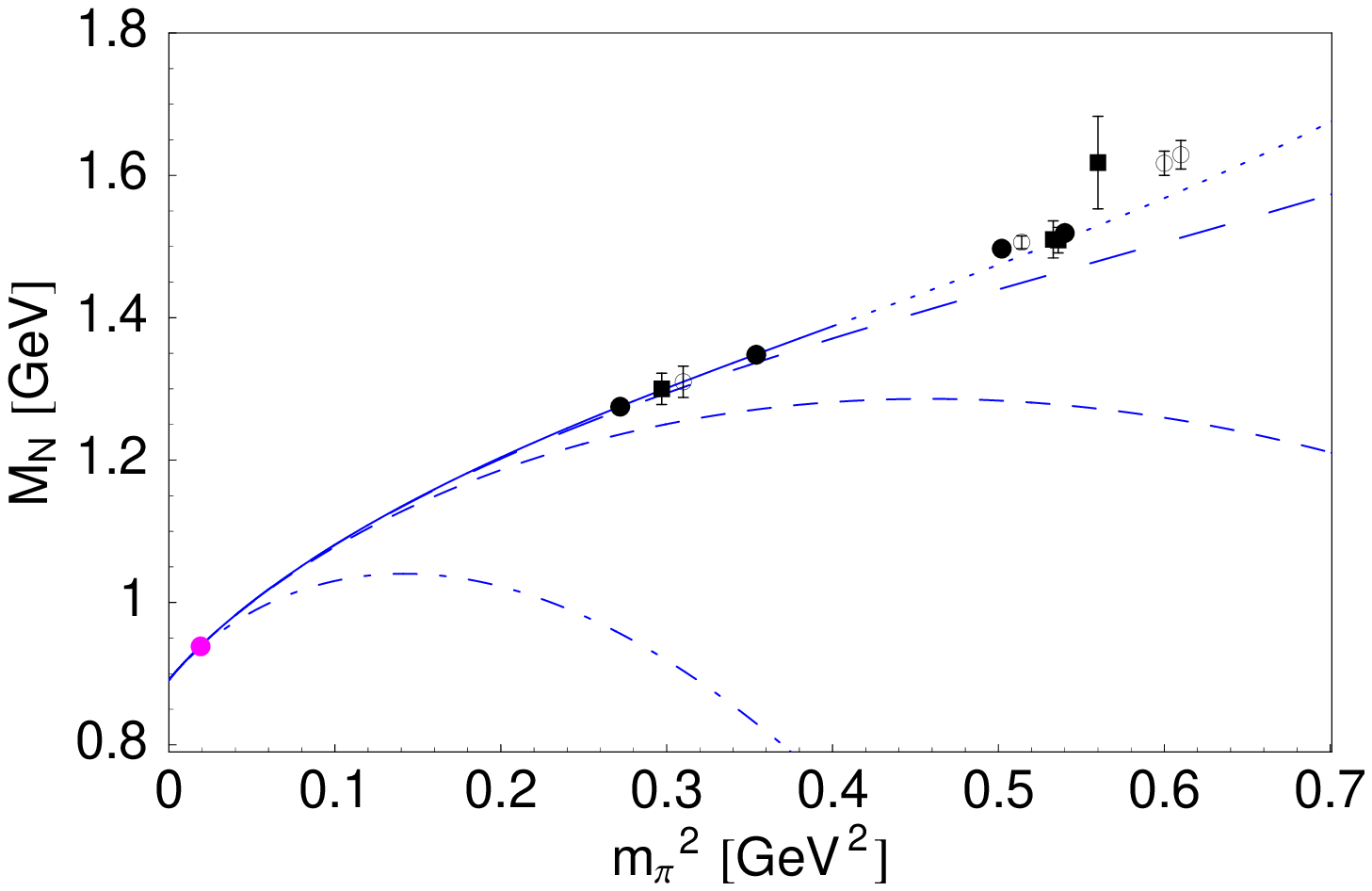}
\caption{Best fit (solid curve) interpolating between lattice results
\cite{CPP,JLQ,QSF} and the physical nucleon mass, using NNLO chiral
perturbation theory \cite {PHW}. The dashed and dash-dotted curves show
consecutive steps in the expansion~\cite{LTY}.}
\label{fig.2}
\end{minipage}
\end{figure}

\subsection{Scalar form factor of the nucleon}

The prominent role played by the pion as a Goldstone boson of spontaneously
broken chiral symmetry has its strong impact on the low-energy structure
and dynamics of nucleons. When probing the individual nucleon with long-wavelength fields, 
a substantial part of the response comes from the pion cloud, the "soft'' surface of the nucleon. While these features are well known and established for the electromagnetic form factors of the nucleon, its scalar-isoscalar meson cloud is less familiar and frequently obscured by the notion of an "effective sigma meson". On the other hand, the scalar field of the nucleon is at the origin of the intermediate range nucleon-nucleon force, the source of attraction that binds nuclei. Let us therefore have a closer look, guided by chiral effective field theory.

Consider the nucleon form factor related to the scalar-isoscalar quark density, $G_S(q^2) = \langle N(p')|\bar{u}u +  \bar{d}d|N(p)\rangle$, at squared momentum transfer $q^2 = (p - p')^2$. In fact, a better quantity to work with is the form factor $\sigma_N(q^2) = m_qG_S(q^2)$ associated with the scale invariant object $m_q(\bar{u}u +  \bar{d}d)$.  Assume that this form factor can be written as a subtracted dispersion relation:
\begin{equation}
\sigma_N(q^2 = -Q^2) = \sigma_N - {Q^2\over\pi}\int_{4m_\pi^2}^\infty dt{\eta_S(t)\over t(t+Q^2)}~~,
\end{equation}
where the sigma term $\sigma_N$ introduced previously enters as a subtraction constant. We are interested in spacelike momentum transfers with $Q^2 = -q^2 \geq 0$. The dispersion integral in Eq.(19) starts out at the two-pion threshold. It involves the spectral function $\eta_S(t)$ which includes all $J^\pi = 0^+, I=0$ excitations coupled to the nucleon: a continuum of even numbers of pions added to and interacting with the nucleon core. 

\begin{figure}[htb]
\begin{minipage}[t]{50mm}
\centering
\includegraphics[scale=0.35,angle=0]{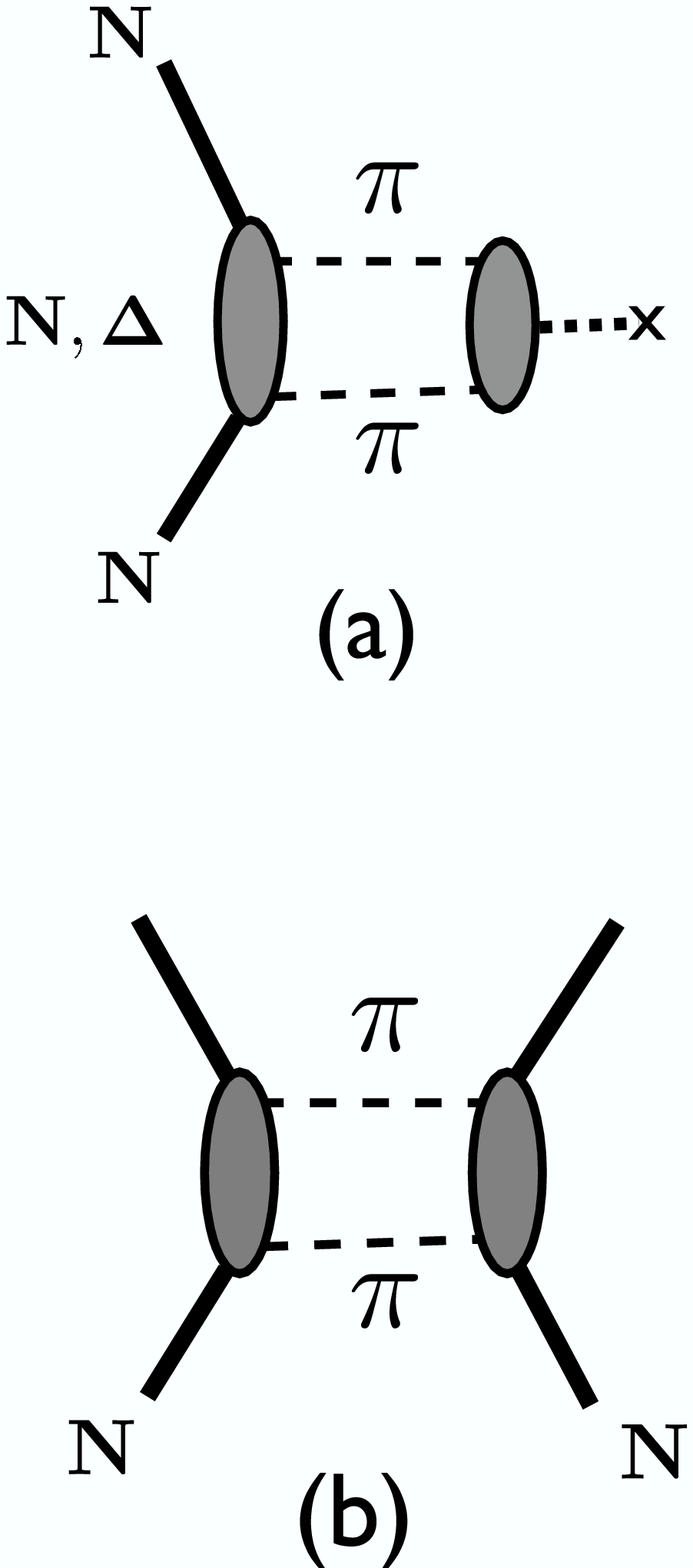}
\caption{(a) Sketch of the scalar form factor of the nucleon; (b) Two-pion exchange interaction between nucleons.}
\label{fig.1b}
\end{minipage}
\hspace{\fill}
\begin{minipage}[t]{90mm}
\includegraphics[scale=0.45,angle=0]{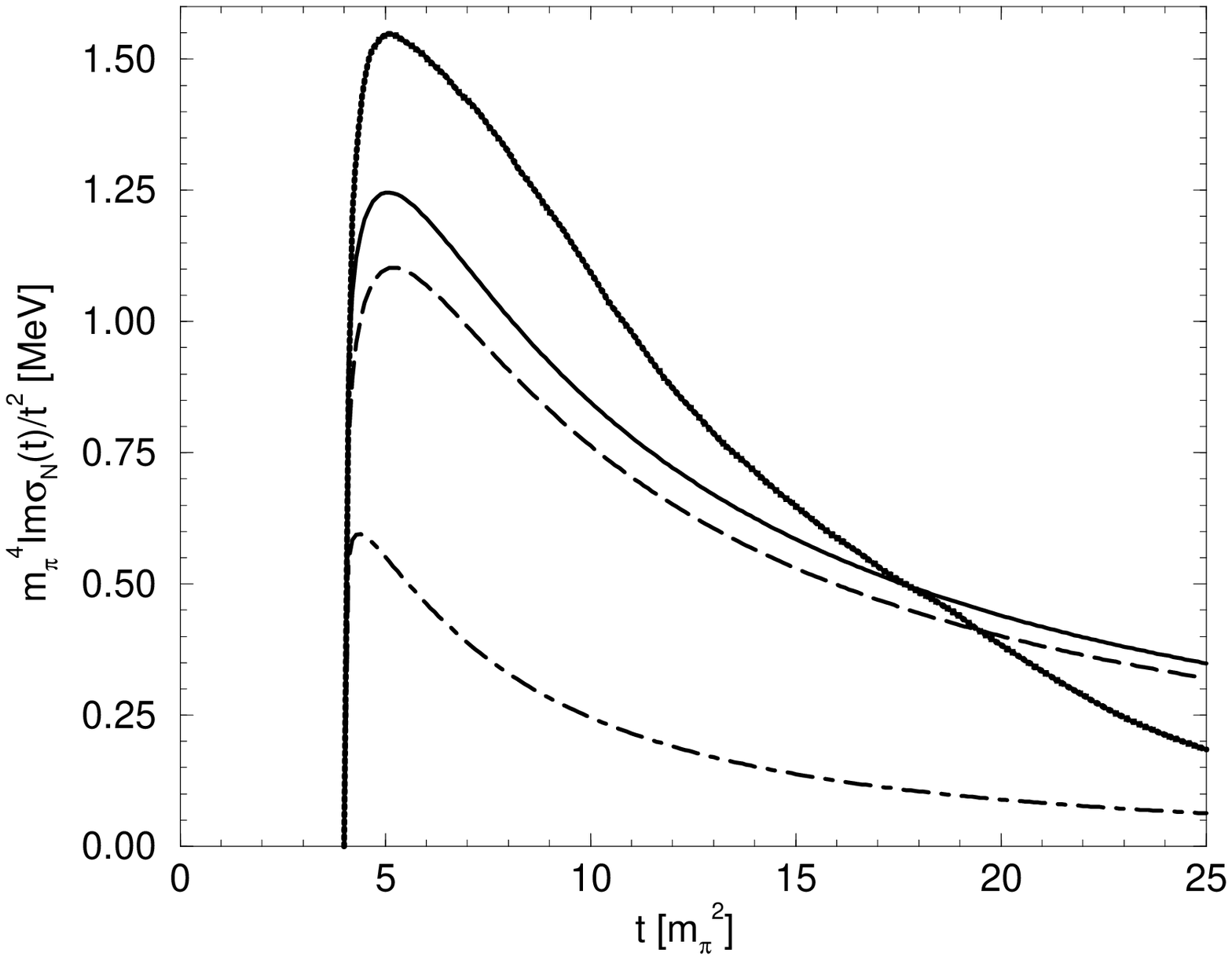}
\caption{Spectral function of the scalar-isoscalar nucleon form factor: chiral
perturbation theory (ChPT at 2 loop NNLO \cite{K}, solid and dashed) in
comparison with empirically deduced spectrum \cite{Hoe} (upper curve). The
separate contribution of the nucleon Born term is also shown (dash-dotted).}
\label{fig.2b}
\end{minipage}
\end{figure}

Chiral perturbation theory at next-to-next-to-leading order (NNLO) in two-loop approximation has been applied in a recent calculation of the spectral function $\eta_S(t)$ \cite{K}. This calculation includes not only nucleon Born terms and leading $\pi\pi$ interactions but also important effects of intermediate $\Delta$ isobar excitations in the two-pion dressing of the nucleon core (see Fig.3a). The result (Fig.4) can be compared with the "empirical" scalar-isoscalar spectral function deduced by analytic continuation from $\pi N$, $\pi\pi$ and $\bar{N}N\leftrightarrow \pi\pi$ amplitudes \cite{Hoe}. Note that there is no such thing as a "sigma meson" in this spectrum which is completely determined by the (chiral) dynamics of the interacting $\pi\pi$ and $\pi N$ system. The integral over $\eta_S(t) t^{-2}$ is proportional to the mean squared scalar radius of the nucleon. One finds \cite{GLS}
\begin{equation}
\langle r_S\rangle^{1/2} \simeq 1.3\,fm~,
\end{equation}
the largest of all nucleon radii, considerably larger than the proton charge radius of 0.86 fm.  

\subsection{Scalar two-pion exchange}

By its magnitude and range, the form factor $G_S(q^2)$ implies that the nucleon, surrounded with its two-pion cloud, is the source of a strong scalar-isoscalar field with a large effective coupling constant $g_S = G_S(q^2 = 0) = \sigma_N / m_q \simeq 10$. When a second nucleon couples to this scalar field (see Fig. 1b), the resulting two-pion exchange $NN$ interaction $V_{2\pi}$ is reminiscent of a Van der Waals force. More than half of the strength of $V_{2\pi}$ is actually governed by the large spin-isospin polarisability of the nucleon related to the transition $N\rightarrow \Delta$ in the intermediate state. At long and intermediate distances it behaves as \cite{KBW}
\begin{equation}
V_{2\pi}(r) \sim {e^{-2m_\pi r}\over r^6}P(m_\pi r)~ ,
\end{equation}
where $P$ is a polynomial in $m_\pi r$. In the chiral limit $(m_\pi \rightarrow 0)$, this $V_{2\pi}$ approaches the characteristic $r^{-6}$ dependence of a non-relativistic Van der Waals potential. 

The two-pion exchange force is the major source of intermediate range attraction that binds nuclei. This is, of course, not a new observation. For example, the important role of the second-order tensor force from iterated
pion exchange had been emphasised long ago \cite{KB}, as well as the close connection of the nuclear force to the strong spin-isospin polarisability of the nucleon \cite{EF}. The new element that has entered the discussion more recently is the systematics provided by chiral effective field theory in dealing with these phenomena.

\section{CHIRAL DYNAMICS AND NUCLEAR MATTER}

\subsection{Scales at work}

In nuclear matter, the relevant momentum scale is the Fermi momentum $k_F$. Around
the empirical saturation point with $k_F^{(0)} \simeq 0.26$ GeV $ \sim 2m_{\pi}$, the Fermi
momentum and the pion mass are scales of comparable magnitude. This implies that at the densities of 
interest in nuclear physics, $\rho \sim \rho_0 = 2(k_F^{(0)})^3/3\pi^2 \simeq 0.16$ fm$^{-3} \simeq 0.45\, m_\pi^3$, pions must be included as {\it explicit} degrees of freedom: their propagation in matter is "resolved" at the relevant momentum scales around the Fermi momentum.

At the same time, $k_F$ and $m_{\pi}$ are small compared to the characteristic chiral scale, $4 \pi f_{\pi} \simeq 1.2$ GeV. Consequently, methods of chiral perturbation theory are expected to be applicable to nuclear matter at least in a certain window around $k_F^{(0)}$. In that range, the energy density
\begin{equation}
{\cal E}(k_F) = \left[M_N + {E(k_F)\over A}\right]\rho\,.
\end{equation}
should then be given as a convergent power series in the Fermi momentum. This is our working hypothesis. More precisely, the energy per particle has an expansion
\begin{equation}
{E(k_F)\over A} = {3k_F^2\over 10 M_N} + \sum_{n\ge3}{\cal F}_n(k_F/m_\pi)\,k_F^n\,.
\end{equation}
The expansion coefficients ${\cal F}_n$ are in general non-trivial functions of $k_F/ m_{\pi}$, the dimensionless ratio of the two relevant scales. These functions must obviously not be further expanded. Apart from $k_F$ and $m_\pi$, a third relevant "small" scale is the mass difference $\delta M = M_\Delta - M_N \simeq 0.3$ GeV between the $\Delta(1232)$ and the nucleon. The strong spin-isospin transition from the nucleon to the $\Delta$ isobar is therefore to be included as an additional important ingredient in nuclear many-body calculations, so that the ${\cal F}_n$ become functions of both $k_F / m_\pi$ and $m_\pi / \delta M$.

Let us get a first impression of how the separation of scales controls the scattering amplitude $T$ for two nucleons (with momenta $|\vec{p}\,| \le k_F$) interacting in the nuclear medium. Omitting its detailed operator structure here for brevity,  we denote this amplitude as $T(Q^2; k_F)$ and assume that it satisfies, at fixed energy, a subtracted dispersion relation  
\begin{equation}
T(Q^2; k_F) = \sum_i \left[T_i^{(0)} + {Q^2\over \pi}\int_{\mu_i^2}^\infty dt {\zeta_i(t; k_F)\over t(t+Q^2)}\right]\, ,
\end{equation}
with subtraction constants $T_i^{(0)}$. Here $Q^2 > 0$ is the (spacelike) squared momentum transfered in the t-channel. The amplitudes (with index $i$) are grouped in a hierarchy of terms according to the number of Goldstone bosons (pions) exchanged between the two nucleons, or any other mechanisms that may occur at short distance (see Fig.5). The corresponding spectral functions $\zeta_i$ start at a characteristic threshold $\mu_i^2$ in each channel (e.g. $\mu_2^2 = 4m_\pi^2$ for two-pion exchange, $\mu_3^2 = 9m_\pi^2$ for three-pion exchange, and so forth). The influence of the Pauli principle in intermediate NN states generally leads to a $k_F$-dependence of these spectral functions.

In the ground state of nuclear matter, the external nucleon lines in Fig.5 have momenta in the Fermi sea. The exchanged momentum is limited to $Q \le 2 k_F$, with an average at $Q \sim k_F$. Clearly, in all those processes which have spectral functions starting at $\mu_i^2 \gg k_F^2$, the amplitudes in (14) reduce to the approximate form $T_i = T_i^{(0)} + d_i^2\, Q^2$ where $d_i^2 = (1/\pi)\int_{\mu_i^2}^\infty dt\, t^{-2}\zeta_i(t)$ is proportional to the mean square distance over which this interaction takes place. These short-distance contributions with $d_i^2\,Q^2 \ll 1$ are thus represented as contact terms (the constants $T_i^{(0)}$), corrected by small finite range (derivative) terms. On the other hand, processes with $\mu_i^2 \le k_F^2$ obviously require a full treatment of their spectral integrals in Eq.(14). For nuclear densities with $k_F^{(0)} \simeq 2 m_\pi$, one- and two-pion exchange processes (as well as those involving low-energy particle-hole excitations) must therefore be treated explicitly. They govern the long-range interactions at the distance scales $d > 1$ fm relevant to the nuclear many-body problem, whereas
short-range mechanisms, with t-channel spectral functions involving much larger masses, are not resolved in detail at nuclear Fermi momentum scales and can be subsumed in contact interactions. This is the "separation of scales" argument that makes strategies of chiral effective field theory work even for nuclear problems, with the "small" scales ($k_F, m_\pi, \delta M$) distinct from the "large" ones ($4\pi f_\pi, M_N$). The filled Fermi sea of nucleons is an important prerequisite for applying low-energy perturbative expansions which would not work in the vacuum.  
\begin{figure}[htb]
\begin{minipage}{70mm}
\includegraphics[scale=0.25,angle=0]{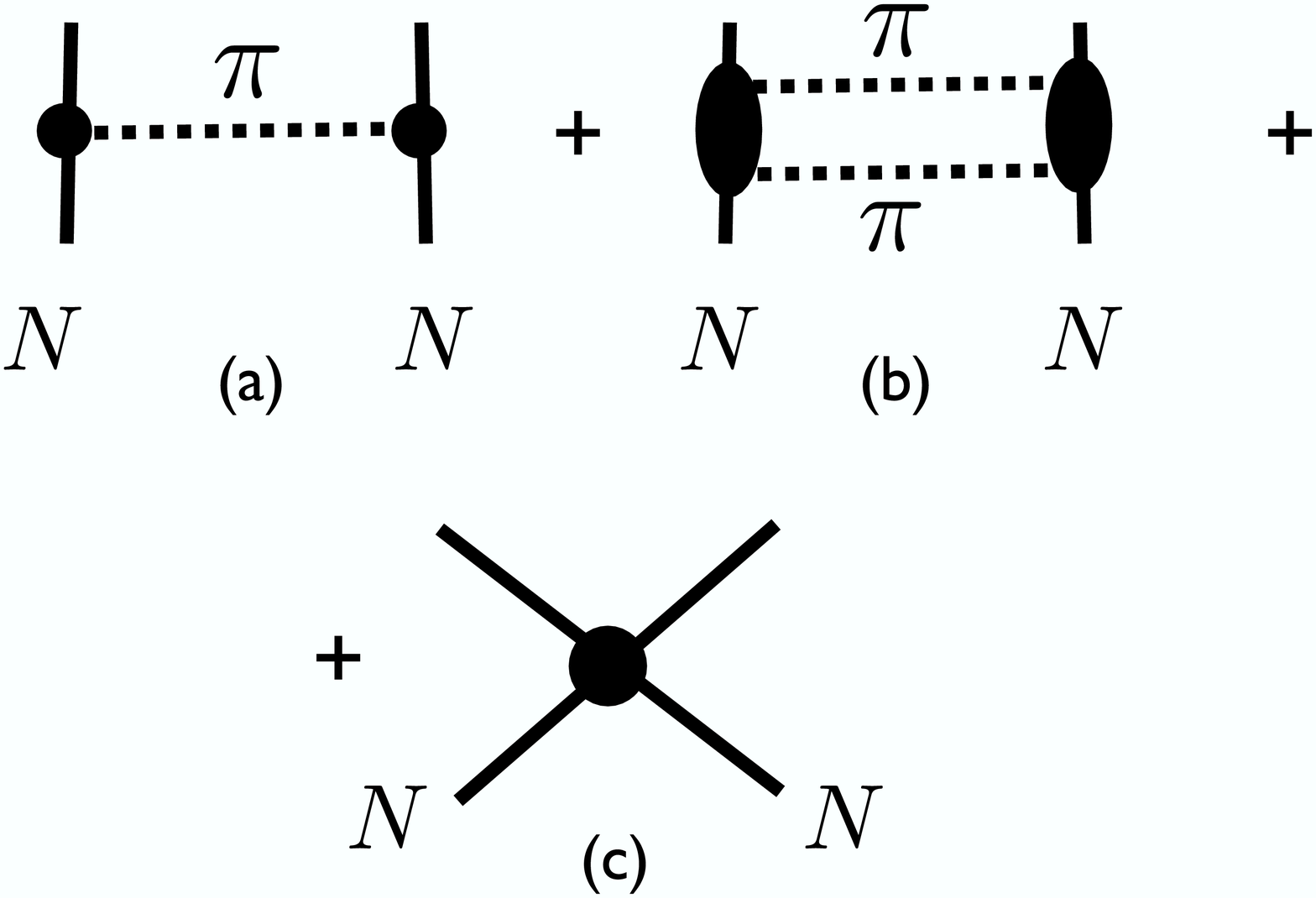}
\caption{NN amplitude in chiral effective field theory: (a) one-pion exchange, (b) two-pion exchange (including $\Delta$ isobar intermediate states), (c) contact terms representing short-distance dynamics.}
\label{fig.5}
\end{minipage}
\hspace{\fill}
\begin{minipage}{65mm}
\includegraphics[scale=0.5,angle=0]{brijuni1.epsi}
\includegraphics[scale=0.5,angle=0]{brijuni2.epsi}
\caption{Energy density from in-medium chiral perturbation theory at
three-loop order. Dashed lines show pions. Each (solid) nucleon line  means
insertion of the in-medium propagator (15). Not shown is an additional
two-loop diagram involving NN contact interactions.}
\label{fig.6}
\end{minipage}
\end{figure}

\subsection{In-medium chiral perturbation theory}

The chiral effective Lagrangian (3) generates the basic pion-nucleon coupling 
terms used in leading-order ChPT. It is of interest  \cite{LFA,KFW} to explore whether 
and to what extent this
leading-order chiral dynamics  can already produce binding and saturation of nuclear matter. The new ingredient in performing calculations at
finite density (as compared to evaluations of scattering processes in vacuum)
is the in-medium nucleon propagator. For a relativistic nucleon with
four-momentum $p^{\mu} = (p_0, \vec{p} \,)$ it reads
\begin{equation}
(\not \!p + M_N) \left\{\frac{i}{p^2 - M_N^2 + i \varepsilon} - 2 \pi \delta (p^2
- M_N^2) \theta (p_0) \theta(k_F - | \vec{p}\, |) \right\}.
\end{equation}
The second term is the medium insertion which accounts for the fact that the
ground state of the system has changed from an "empty" vacuum to a filled Fermi
sea of nucleons. Diagrams can then be organized systematically  in the number
of medium insertions, and an expansion is performed in leading inverse powers
of the nucleon mass, consistently with the $k_F$-expansion.

Our "inward-bound" strategy \cite{KFW} is now as follows. One starts at large
distances\footnote{Note however that this scheme cannot be pursued down to extremely low densities where new non-perturbative phenomena begin to take over, such as formation of nucleon clusters.} (small $k_F$) and systematically generates the pion-induced
correlations between nucleons as they develop with decreasing distance
(increasing $k_F$). Calculations along these lines \cite{KFW} have been 
performed to 3-loop order in the energy density (see Fig.6),
including terms up to order $k^5_F$. These calculations incorporate one- and two-pion
exchange processes. Hartree contributions from one-pion exchange vanish for a spin-saturated system, and the one-pion exchange Fock term (upper left of Fig.6) is small, so the leading effect comes from the exchange of two pions, with the second order tensor force (in the upper middle diagram of Fig.6) providing the dominant attraction. 

At order $k_F^3$, the momentum space loop integral with iterated 
pion exchange encounters a divergence which needs to be regularised. This can be done in two equivalent ways. Either one introduces a high-momentum cutoff $\Lambda$; or one removes the divergent parts using  dimensional regularisation and replaces them by a counter term. Both procedures are equivalent to introducing a subtraction constant in a corresponding dispersion relation approach. In either case one ends up with an adjustable $NN$ contact interaction  which encodes dynamics at short distances not resolved explicitly in the effective low-energy theory. This single contact term is the only free
parameter at this stage. Adjusting it to the binding energy of nuclear matter, the outcome is quite remarkable \cite{KFW}: despite its simplicity, this schematic one-parameter approach already produces nuclear binding and saturation (see Fig.7). One finds $E(\rho_0)/A = -15.3$ MeV at an equilibrium density $\rho_0 = 0.178$ fm$^3$ together with a (predicted) compressibility $K = 255$ MeV (empirical: $K = 220 \pm 50$ MeV \cite{B}). The extension to asymmetric nuclear matter gives a calculated asymmetry energy per nucleon, ${\cal A} = 33.8$ MeV with no additional parameter (empirical: ${\cal A} \simeq 33$ MeV \cite{SH}).
This result suggests that isospin dependent interactions are to large extent already accounted for by chiral pion dynamics, without necessity of introducing extra ingredients such as the $\rho$ meson. 

In the scenario just described, entirely based on just leading order chiral pion-nucleon dynamics, the binding and saturation of nuclear matter comes from the balance between two mechanisms: (a) second order pion exchange with high-momentum cutoff which gives driving attraction proportional to $k_F^3 \sim \rho$ in the energy per particle; (b) the Pauli exclusion principle acting on intermediate nucleon states of the two-pion pion exchange processes, which generates a repulsive contribution of order $k_F^4$. This ${\cal O}(k_F^4)$ term is finite and uniquely determined in a model independent way, free of adjustable parameters, involving only known low-energy constants ($g_A, f_\pi$) and masses ($m_\pi, M_N$). This statement remains unchanged when higher orders in the small-scale expansion are included. 

It is instructive to examine the expansion (13) in the chiral limit ($m_\pi \rightarrow 0$)
where the basic saturation mechanism is already apparent \cite{KFW}. In this limit the coefficients of the $k_F^3$ and $k_F^4$ terms of $E/A$ become:
\begin{equation}
{\cal F}_3 = -10M_N\Lambda \left({g_A\over 4\pi f_\pi}\right)^4 \, , ~~~~~~ {\cal F}_4 = {3M_N\over 70}\left({g_A\over 4\pi f_\pi}\right)^4[4\pi^2 + 237 - 24\,ln\, 2] ~~,
\end{equation}
where small corrections from the one-pion exchange Fock term and from the expansion of the kinetic energy have been dropped. One notes that 
the model-independent ${\cal F}_4 = 0.115$ fm$^{-3}$ is remarkably close to the empirical ${\cal F}_4 \simeq 0.11$ fm$^{-3}$. A high-momentum cutoff scale $\Lambda$ in the 0.5 GeV range matches the equivalent contact interaction proportional to $\rho$ for which the empirically required strength is ${\cal F}_3 \simeq -0.234$ fm$^{-2}$.
\begin{figure}[htb]
\begin{minipage}{70mm}
\includegraphics[scale=0.31,angle=0]{fig7.eps}
\caption{Energy per particle of symmetric nuclear matter as determined by a one-parameter in-medium ChPT approach \cite{KFW} (solid curve). The dashed curve shows for comparison an example of a sophisticated many-body calculation \cite{FP}.}
\label{fig.7}
\end{minipage}
\hspace{\fill}
\begin{minipage}{85mm}
\quad
\includegraphics[scale=0.27,angle=-90]{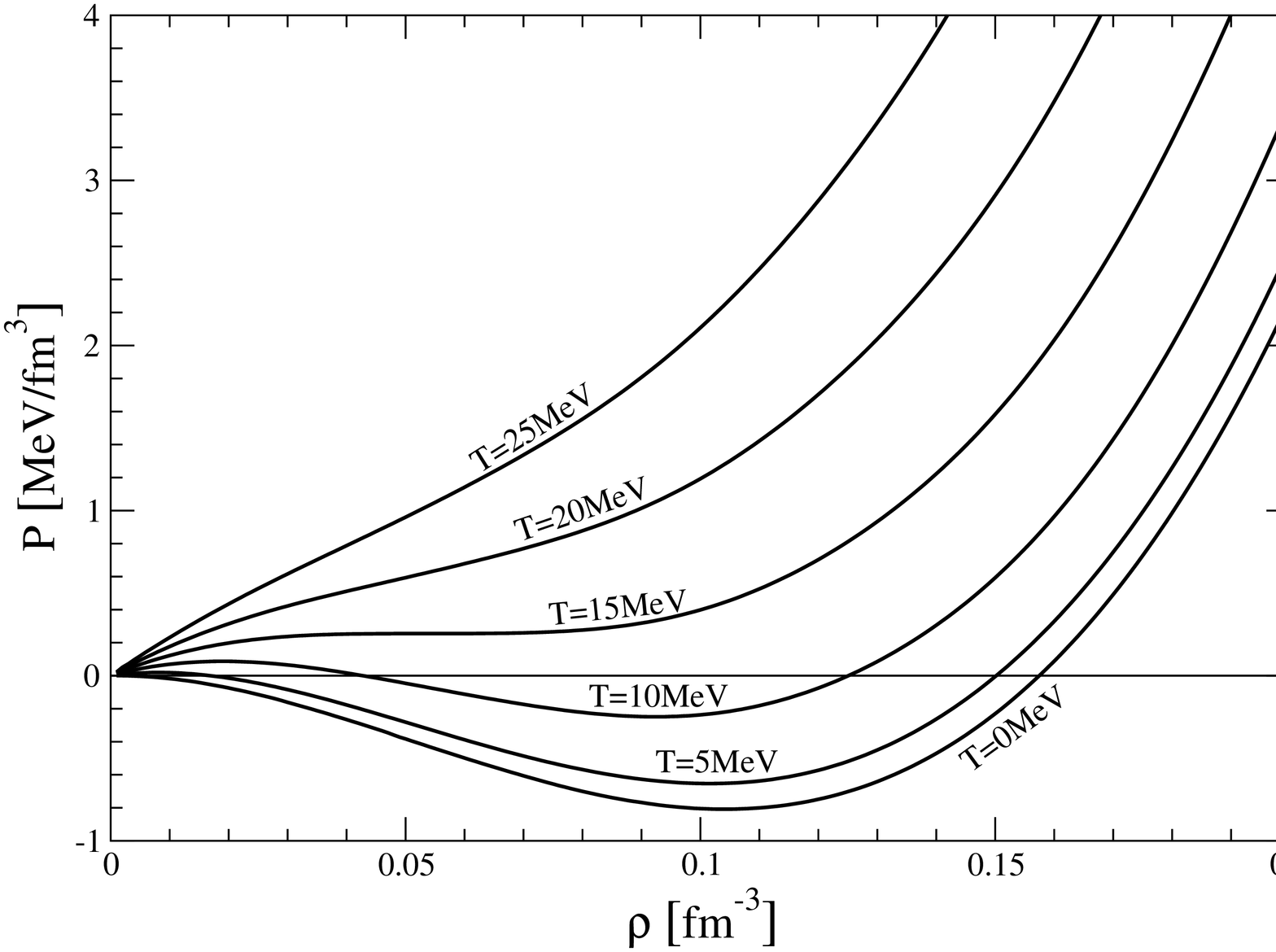}
\caption{Equation of state of symmetric nuclear matter, showing the isotherms of the pressure $P(\rho, T)$ as function of baryon density. The calculation \cite{FKW4} includes pions, nucleons and $\Delta$ isobars at three-loop order in the free-energy density, plus corrections from three-body terms at ${\cal O}(k_F^6)$ in the energy per particle. }
\label{fig.8}
\end{minipage}
\end{figure}

While these exploratory first steps look quite promising, reproducing bulk properties of nuclear (and neutron) matter is just a necessary but by no means sufficient condition to deal with the nuclear many-body problem. Chiral dynamics with just pions and nucleons gives a surprisingly good start for first orientation, but at that (yet incomplete) level of the discussion, details such as the momentum dependence of the nucleon single particle potential and the density of states at the Fermi surface do not yet come out satisfactorily \cite{FKW2}. A further indicator for still missing degrees of freedom is the fact that chiral dynamics restricted to pions and nucleons alone gives the critical temperature for the nuclear liquid-gas phase transition \cite{FKW3} too large by about 20-25\%. 

The next important step is to incorporate the strong $N\rightarrow\Delta$ transition in the two-pion exchange dynamics, together with related three-nucleon forces and Pauli corrections. Such effects were also found to be important in the ab-initio calculations of the Illinois group \cite{PPWC}. This extended chiral approach \cite{FKW4} has two subtraction constants encoding unresolved short-distance dynamics in the functions ${\cal F}_{3,5}(k_F/m_\pi; m_\pi / \delta M)$ and a genuine three-body term which corrects ${\cal F}_6$, but no changes appear in the ${\cal F}_4$ term from the inclusion of $\Delta$ isobars. The results for both nuclear and neutron matter are now much improved. In particular, the momentum dependent (complex) nucleon single-particle potential comes out very well in the momentum range $p < 400$ MeV/c. The Hugenholz - Van Hove theorem is strictly fulfilled throughout the procedure. 

In this scenario, nuclear many-body dynamics emerges primarily from the Van der Waals - like nature of the attractive two-pion exchange forces in combination with the repulsive action of the Pauli principle. It is therefore perhaps not surprising that the resulting equation of state at finite temperature \cite{FKW4} (see Fig.8) is reminiscent of the one familiar from a Van der Waals gas. The predicted liquid-gas transition temperature $T_c \simeq 15$ MeV is now close to the empirically deduced value, $T_c = 16.6 \pm 0.9$ MeV \cite{N}. 

At that stage the present approach, guided by chiral effective field theory representing low-energy QCD, treats nucleons basically as non-relativistic particles: the low-momentum, small-scale expansion of the energy density is simultaneously an expansion in inverse powers of the nucleon mass. Contact with the phenomenological non-relativistic Skyrme-type energy density functional can be made at this point \cite{FKW4,KFW2}. General features of nuclear Skyrme phenomenology - including surface (gradient) terms - are well reproduced around the densities relevant for nuclei (although the chiral approach predicts pronounced density dependence of the Skyrme parameters at low densities, induced by strong pionic effects when the Fermi momentum becomes comparable to the pion mass).

 A further interesting comparison can be made with the universal low-momentum NN interaction, $V$(low-k), derived from a variety of phase-shift equivalent potentials using renormalisation group methods \cite{BKS}. At the Hartree level (linear in the density $\rho$), the single particle potential generated by $V$(low-k) agrees well with the one derived from chiral dynamics, whereas non-trivial density dependence takes over in the chiral potential already at relatively low density, reflecting the action of the Pauli principle on in-medium two-pion exchange processes \cite{FKW4}.

There is one place, however, where the non-relativistically reduced chiral theory fails completely: it misses the strength of the nuclear spin-orbit force by a large factor. The detailed investigation of mechanisms which generate a spin-orbit interaction from two-pion exchange (both from the second order tensor force and from terms with intermediate $\Delta$'s) reveals that these contributions cancel to a large extent \cite{FKW4}. This is a hint that genuine relativistic effects, not visible in bulk properties of infinite homogeneous nuclear matter, must be considered. 
  
\subsection{Mean fields from QCD condensates}

In relativistic nuclear models \cite{SW,RF}, strong Lorentz scalar and vector mean fields, each several hundred MeV in magnitude, are at the origin of the abnormally large spin-orbit splitting observed in nuclei. While these scalar and vector fields roughly cancel in the average single particle potential so that their individual strengths are not revealed in the energy per nucleon, they act coherently in building up the spin-orbit potential. In-medium QCD sum rules \cite{CFG,DL} give some guidance as to where such strong fields have their sources.  

Consider the scalar and vector self-energies $\Sigma_{S,V}(\rho)$ appearing in the Dirac equation
$[\gamma_\mu(i\partial^\mu - \Sigma_V^\mu) - (M_N + \Sigma_S)]\Psi_N = 0$ of a nucleon which propagates in isospin-symmetric nuclear matter. QCD sum rules establish a relationship between the leading terms of $\Sigma_S$ and $\Sigma_V \equiv \Sigma_V^{\mu = 0}$, and the changes with density of the lowest-dimensional quark condensates, $\langle \bar{q} q \rangle$ and $\langle \bar{q} \gamma_0 q \rangle = \langle q^\dagger q\rangle = 3\rho/2$.
In leading order the condensate 
part of $\Sigma_S$ is expressed in terms of the density dependent 
chiral condensate as follows \cite{CFG,DL}: 
\begin{equation} 
\Sigma_S= - \frac{8 \pi^2}{\Lambda_B^2} [ 
\langle \bar{q} q \rangle_\rho - \langle \bar{q} q \rangle_0 ] 
 = - \frac{8 \pi^2}{\Lambda_B^2}~\frac{\sigma_N}{m_u +m_d} \rho_S = -{\sigma_N M_N \over m_\pi^2 f_\pi^2}\rho_S \; ,
\end{equation} 
with the nucleon scalar density $\rho_S = \langle \bar{\Psi}_N\Psi_N \rangle$. The difference between the vacuum condensate $\langle \bar{q} q \rangle_0$ 
and the one at finite density involves the nucleon sigma term (5).
The Borel scale $\Lambda_B$ which roughly separates 
perturbative and non-perturbative domains in the QCD sum rule 
analysis, has been eliminated in the last step, using the Ioffe relation (7) together with $(m_u + m_d)\langle \bar{q}q \rangle = -m_\pi^2f_\pi^2$. The interpretation of Eq.(17) is as follows: once nucleons are present in the QCD vacuum, they polarise this vacuum in such a way that the magnitude of the (negative) chiral condensate gets reduced with increasing density. The resulting attractive scalar mean field is large:
\begin{equation}
\Sigma_S = M_N^*(\rho) - M_N \simeq -6.9\, \sigma_N{\rho_S\over\rho_0}\,,
\end{equation}
or $\Sigma_S \simeq -350$ MeV at $\rho_0 = 0.16$ fm$^{-3}$ (using $\sigma_N \simeq 50$ MeV), implying that the nucleon mass in nuclear matter is effectively reduced by more than 1/3 of its vacuum value.  To the same order in the condensates with lowest dimension, the time component of the isoscalar vector self-energy is 
\begin{equation} 
\Sigma_V = \frac{64 \pi^2}{3 \Lambda_B^2} \langle q^\dagger q \rangle_\rho 
= \frac{32 \pi^2}{\Lambda_B^2} \rho = {4(m_u + m_d) M_N \over m_\pi^2 f_\pi^2} \rho \; .
\end{equation}  
It reflects the repulsive density-density correlations associated with the time component
of the quark current, $\bar{q}\gamma^{\mu}q$. Note that, as pointed out in ref.\cite{CFG}, the ratio 
\begin{equation} 
\label{ratio} 
\frac{\Sigma_S}{\Sigma_V} = - \frac{\sigma_N}{4(m_u +m_d)} 
\frac{\rho_S}{\rho} 
\end{equation} 
is approximately equal to  $-1$ for typical values of 
the nucleon sigma term $\sigma_N$ and the current quark masses 
$m_{u,d}$, and around nuclear matter saturation density where $\rho_S \simeq \rho$ 
(as an example, take $\sigma_N \simeq  50$ MeV and $m_u +m_d \simeq  
12$ MeV at a renormalisation scale  corresponding to the chiral "gap" $4\pi f_\pi \sim 1$ GeV). 
The individually strong scalar and vector fields thus balance each other in the average single particle potential, $U \simeq \Sigma_S + \Sigma_V$, and the resulting small contribution simply gets absorbed in the subtraction constant attached to the $k_F^3$ term of the chiral expansion of the energy per nucleon. However, when introducing local densities for finite systems so that $\Sigma_S(\rho_S(r)) \equiv S(r)$ and $\Sigma_V(\rho(r)) \equiv V(r)$, the spin-orbit potential
\begin{equation}
U_{s.o.} = {1\over 2M_N^2 r}\,{d\over dr}\left({V-S\over 1-{V-S\over 2M_N}}\right)\vec{l}\cdot\vec{s}
\end{equation}
has a huge strength proportional to $V(0)-S(0) \simeq 0.7$ GeV, just the one required by phenomenology.

While useful for further orientation, the QCD sum rule constraints (17-20) are of course not very precise at a quantitative level. The estimated error in the ratio $\Sigma_S/\Sigma_V\simeq -1$ 
is about 20\%, given the limited accuracy in the values of $\sigma_N$ and $m_u+m_d$. 
The leading-order Ioffe formula (7) on which Eq.(17) relies has corrections 
from condensates of higher dimension. In previous QCD sum rule studies the largest uncertainty came from the unknown density dependence of the four-quark condensates $\langle \bar{q}q\bar{q}q\rangle$. This uncertainty is now considerably reduced, however, by the explicit calculation of chiral two-pion exchange contributions to the in-medium nucleon self-energy. 

\section{FINITE NUCLEI}

The translation to finite nuclei is best performed using the (relativistic) density functional framework (for recent reviews, see ref.\cite{LRV}). Schematically, the energy of the nucleus is written as
\begin{equation}
E[\rho] = T_{kin} + \int d^3x \,[{\cal E}_{bg}(\rho) + {\cal E}_{ex}(\rho)] + E_{coul}\,.
\end{equation}
with the density $\rho = \sum_{k=1}^{A}|\psi_k\rangle\langle\psi_k|$ expressed in terms of the self-consistent nucleon wave functions $\psi_k$ of the occupied orbits. We associate the "background" part ${\cal E}_{bg}$ of the energy density with the strong scalar and vector mean fields generated by in-medium changes of QCD condensates, while the "exchange correlation" part ${\cal E}_{ex}$ includes the pionic fluctuations and Pauli principle effects, calculated using in-medium chiral perturbation theory as described in section 3.2. At a practical level, the calculations generating the energy density functional are conveniently done by introducing an equivalent effective Lagrangian with density dependent four-point couplings between nucleons,
\begin{equation}
{\cal L}_{eff} = \bar{\Psi}(i\gamma\cdot\partial - M_N)\Psi - {1\over 2}\sum_i G_i(\rho)\left(\bar{\Psi} \Gamma_i\Psi\right)^2  - {1\over 2}\sum_i D_i(\rho)\left(\partial \bar{\Psi} \Gamma_i\Psi\right)^2 + {\cal L}_{e.m.} \, .
\end{equation}
While this auxiliary Lagrangian is formally designed to be used in the mean-field limit, its physics
content reaches out much further. Quantum fluctuations beyond mean field are incorporated through the density dependent coupling strengths, $G_i(\rho)$ and $D_i(\rho)$, which multiply the interaction terms $(i = S,V, ...)$ with operators $\Gamma_S = {\bf 1}$, $\Gamma_V = \gamma^\mu$ etc. The matching to the nuclear matter calculations is done at the level of the nucleon self-energies expanded in powers of the Fermi momentum, or equivalently, in powers of $\rho^{1/3}$. This procedure involves the so-called rearrangement terms that occur, as a consequence of the density dependent couplings, in order to maintain thermodynamic consistency. Electromagnetic interactions $({\cal L}_{e.m.})$ are also included. 

\begin{figure}[htb]
\begin{minipage}[t]{70mm}
\includegraphics[scale=0.3,angle=-90]{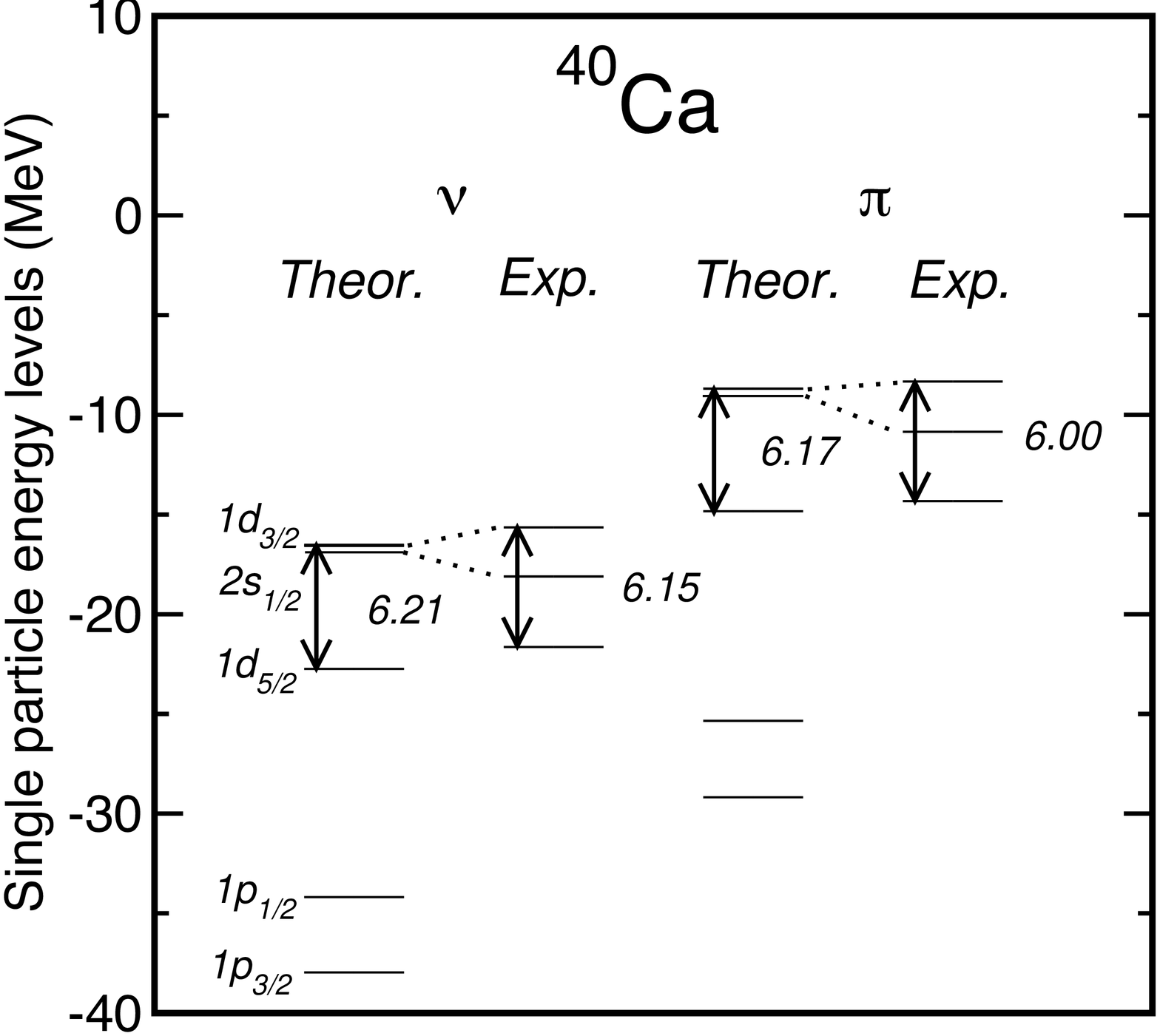}
\caption{Single particle levels for neutrons (left) and protons (right) in $^{40}$Ca in comparison with levels deduced from experiment. The calculations are performed in the relativistic point coupling model constrained by QCD and chiral symmetry \cite{FKVW2}.}
\label{fig.9}
\end{minipage}
\hspace{\fill}
\begin{minipage}[t]{80mm}
\includegraphics[scale=0.3,angle=-90]{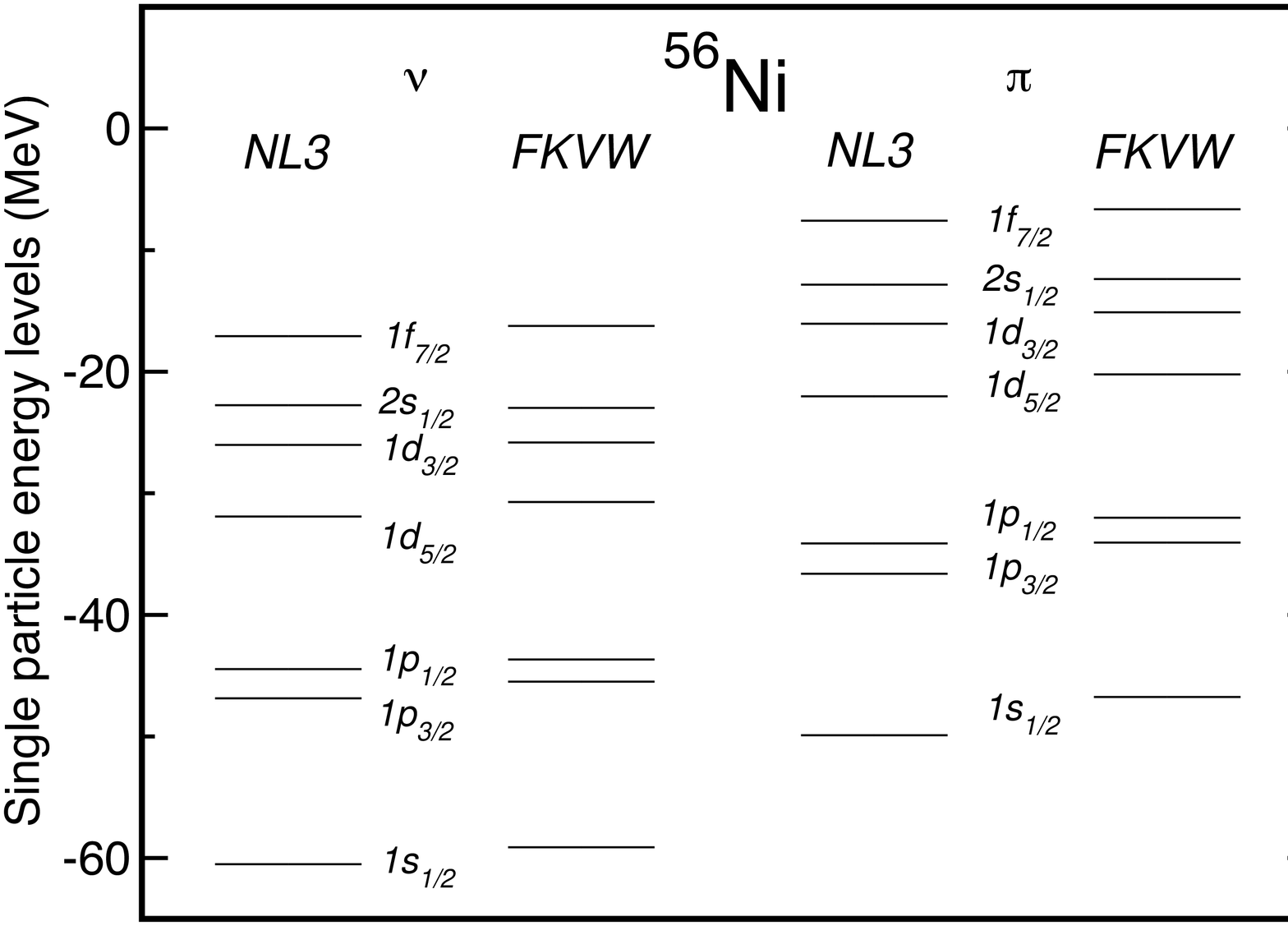}
\vspace{.6cm}
\caption{Single particle levels for protons (left) and neutrons (right) in $^{56}$Ni. Calculations using Eq.(23) with constraints from QCD sum rules and chiral pion dynamics (FKVW \cite{FKVW2}) are compared with the relativistic mean field model (NL3) of ref.\cite{LKR}.}
\label{fig.10}
\end{minipage}
\end{figure}

Detailed calculations along these lines are reported in \cite{FKVW1,FKVW2}, and the framework is summarised and discussed in \cite{VW}. The self-consistent Dirac equation is solved for the single-particle orbits of each given nucleus, so far covering a broad range from $^{16}$O to $^{208}$Pb and to be further extended. The optimal reproduction of binding energies, radii and form factors of the whole set of nuclei requires fine-tuning of the density dependent coupling strengths $G_i(\rho)$. It is remarkable, however, that this fine-tuning systematically remains within less than 10\% of the constraints set by low-energy QCD in terms of chiral pion dynamics and density dependent condensates. The additional surface (derivative) term in Eq.(23) turns out to be perfectly consistent with the corresponding gradient terms found in the in-medium ChPT calculation of inhomogeneous nuclear matter \cite{KFW2,FKW4}. 

We now present a few representative examples of results from such calculations \cite{FKVW2}. Fig.9 shows the single particle spectra for neutrons and protons in $^{40}Ca$. While the binding of these levels is governed by chiral two-pion exchange in the presence of the filled Fermi sea (the "Van der Waals plus Pauli" mechanisms that we referred to earlier), the spin-orbit splitting is completely determined by the strong scalar and vector mean fields that can be interpreted as arising from in-medium changes of the condensate structure of the QCD vacuum. 
\begin{figure}[htb]
\begin{minipage}[t]{150mm}
\includegraphics[scale=0.55,angle=-90]{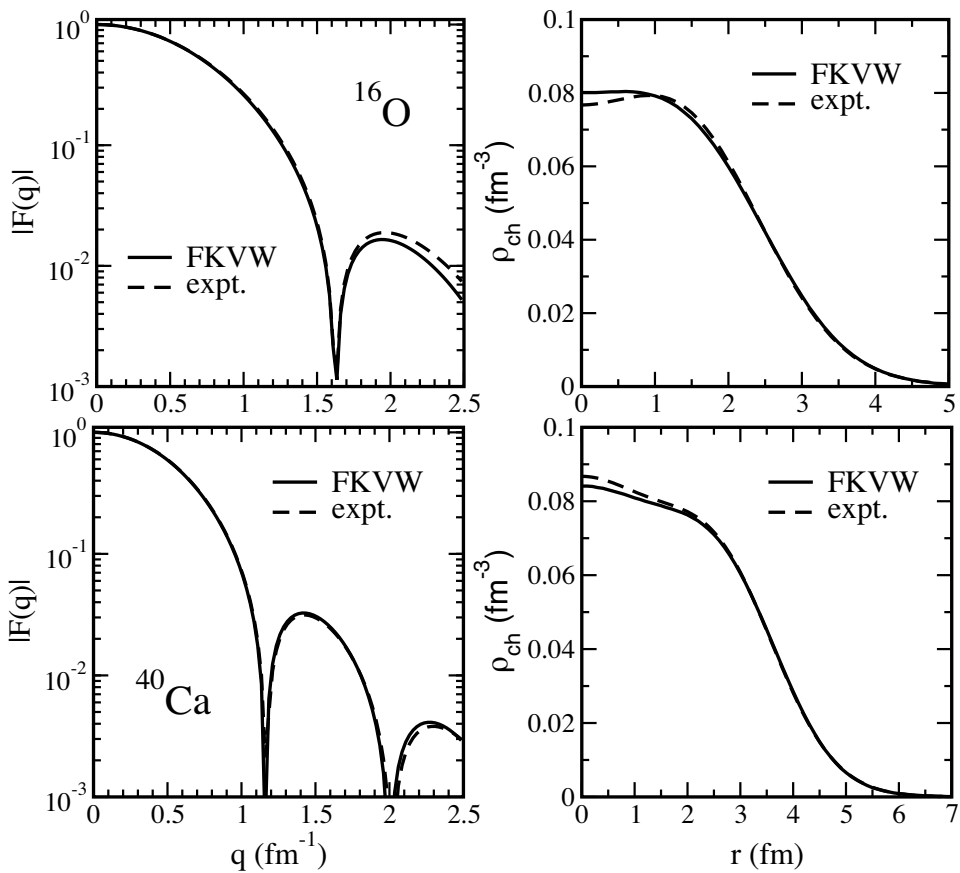}
\caption{Charge density distributions and form factors of $^{40}$Ca and $^{56}$Ni calculated in the relativistic point coupling model (23) of ref.\cite{FKVW2} (solid lines). The dashed lines are the empirical form factors and charge densities deduced from electron scattering experiments.}
\label{fig.5b}
\end{minipage}
\end{figure}
Fig.10 shows our calculations \cite{FKVW2} for $^{56}$Ni in comparison with results from ref.\cite{LKR} chosen as a representative of the currently most advanced relativistic mean field phenomenolgy. Examples of charge distributions and form factors are displayed in Fig.11. These calculations include center-of-mass corrections and convolution with the empirical proton form factor. The direct comparison with experimental data is evidently quite successful, considering that the detailed surface structure of these distributions is reproduced using the gradient terms derived from chiral pion dynamics, with no additional fine-tuning required. 

It is instructive to summarise the trends in the equivalent scalar and vector self-energies for symmetric nuclear matter, reconstructed using the optimised coupling strengths $G_i(\rho)$ found in the overall "best fit" to finite nuclei. These self-energies have the following approximate pattern (with $\rho_0 = 0.16$ fm$^{-3}$):
\begin{equation}
\Sigma_{S,V}(\rho)\, \simeq \,\Sigma_{S,V}^{(0)}\left({\rho\over\rho_0}\right) - 75\,MeV\left({\rho\over\rho_0}\right)\left[1 - 0.6\left({\rho\over\rho_0}\right)^{1/3} - 0.2\, \left({\rho\over\rho_0}\right)^{2/3}  + 0.1\,\left({\rho\over\rho_0}\right)\right] 
\end{equation} 
where $\Sigma_S^{(0)} \simeq -0.34$ GeV and $\Sigma_V^{(0)} \simeq +0.34$ GeV just balance to zero in their summed contribution to the single particle potential but act coherently in  producing the large spin-orbit splitting. These pieces are compatible with the scalar and vector background fields derived from in-medium QCD sum rules. The additional part linear in the density $\rho$ results from contact terms (subtraction constants) in the in-medium ChPT calculation. The terms proportional to $\rho^{4/3}$ and $\rho^{5/3}$ are directly deduced from chiral pion dynamics without tuning, while the correction of order $\rho^2$, representing genuine three-body effects and higher order contributions, is subject to adjustment.

Although the expansion (24) is perturbative by construction, it is a non-trivial feature that the detailed density dependence of the nucleon scalar and vector self-energies as obtained in ref.\cite{FKVW2} follows very closely, over a wide range of densities, the results of Dirac-Brueckner G-matrix calculations \cite{GFF} which start from a realistic boson-exchange NN interaction. The reasoning behind this observation can presumably be traced to the separation of scales at work, a key element of chiral effective field theory which motivates the present approach. One- and two-pion exchange in-medium dynamics at momentum scales around $k_F$ are treated explicitly and include all terms (ladders and others) to three-loop order in the energy density. The non-trivial $k_F$ dependence in the self-energies (beyond "trivial" order $k_F^3$) reflects in large part the action of the Pauli principle in these processes. Such features are also present in the Brueckner ladder summation which includes, for example, Pauli blocking effects on iterated one-pion exchange. These model-independent terms produce the characteristic $k_F^4$ (or equivalently, $\rho^{4/3}$) behaviour in the energy per particle. On the other hand, the iteration of the short-distance pieces of the NN potential to all orders in the Brueckner ladder, involves intermediate momenta much larger than $k_F$ and gets absorbed in contact terms which produce self-energy pieces linear in the density. Note that these contact interactions must {\it not} be iterated further since they already represent the full short-distance T-matrix information. 

\section{SUMMARY AND OUTLOOK} 

The aim of this presentation has been to demonstrate the key role in the structure of nucleons and nuclei played by a guiding principle that governs low-energy QCD: spontaneous chiral symmetry breaking. The effective field theory based on this principle ties together a wide range of strong-interaction phenomena involving the lightest ($u-$ and $d-$)quarks: from the low-energy interactions of pions and nucleons via the condensate structure of the QCD vacuum to basic aspects of nuclear binding and saturation.

We are on the way to a relativistic nuclear energy density functional constrained by low-energy QCD. At the least, these constraints significantly reduce the freedom in the choice of parameters. In this approach, nuclear binding and saturation arises from the Van der Waals - nature of chiral two-pion exchange forces combined with the Pauli principle. Strong scalar and vector mean fields, expected to emerge from in-medium changes of QCD condensates, largely cancel in the energy per nucleon but drive the spin-orbit splitting in finite nuclei.

The genuine isospin dependence of interactions produced by chiral pion-nucleon dynamics offers interesting perspectives for extrapolations to nuclear systems with extreme isospins. A further step,
presently taken \cite{KW}, is the extension of this framework to chiral SU(3) dynamics and its application to hypernuclei.\\

{\bf Acknowledgements:} Special thanks go to Norbert Kaiser whose work is the basis for a large part of the developments reported here. Important contributions by Paolo Finelli, Stefan Fritsch, Thomas Hemmert, Massimiliano Procura and Dario Vretenar to this joint venture are also gratefully acknowledged.

\end{document}